# Real-time full-field imaging through scattering media by all-optical feedback

Ronen Chriki[1], Simon Mahler[1], Chene Tradonsky[1], Asher A. Friesem[1] and Nir Davidson[1*]

[1] Department of Physics of Complex Systems, Weizmann Institute of Science, Rehovot 7610001, Israel

* Corresponding author: nir.davidson@weizmann.ac.il

**Abstract**

Full-field imaging through scattering media is fraught with many challenges. Despite many achievements in recent years, current imaging methods are too slow to deal with fast dynamics that occur for example in biomedical imaging. Here we present an ultra-fast all-optical method, where the object to be imaged and the scattering medium (diffuser) are inserted into a highly multimode self-imaging laser cavity. We show that the intra-cavity laser light from the object is mainly focused onto specific regions of the scattering medium where the phase variations are low. Thus, round trip loss within the laser cavity is minimized, thereby overcoming most of the scattering effects. The method is exploited to image objects through scattering media whose diffusion angle is lower than the numerical aperture of the laser cavity. As our method is based on optical feedback inside a laser cavity, it can deal with temporal variations that occur on timescales as short as several cavity round trips, with an upper bound of 200 ns.

Optical imaging through scattering media, e.g. biomedical imaging and imaging through atmospheric turbulence, is fraught with many difficulties. The complex structure of such media causes random variations in refractive index, such that an incoming beam is scattered and distorted. As a result, images are blurred or speckled when observed through scattering media[1,2]. Over the past decade, several interesting methods have been developed to overcome these deleterious effects, mainly by exploiting spatial correlations of the seemingly random speckle pattern[3–6] or by shaping the wavefront of the incoming beam[7–9]. But methods based on spatial correlations are limited to thin objects, and only apply to small fields of view that are contained within the 'memory effect'[4]. Furthermore, these methods are mostly applicable to fluorescent objects, which typically have weak signals, and therefore require long integration time to obtain sufficient signal to noise ratio. Thus, they are inadequate for real-time imaging of live tissues, which require short exposure times. Methods based on wavefront shaping involve deformable mirrors or spatial light modulators that control the shape of the incoming wavefront, in order to compensate for the effect of the scattering media. These methods usually require a bright reference 'guidestar', and are based on iterative time-consuming computer algorithms, typically lasting several seconds or more. During the extensive time period needed to shape the incoming wavefront, the sample cannot evolve, and must remain steady on a scale smaller than a wavelength. Although these methods can provide a high signal to noise ratio (SNR) for an image of a single pixel, the SNR decreases linearly as the number of pixels increases[8]. These methods are also based on the 'memory effect', and therefore suffer from a limited field of view. Recently, we presented a new method for rapid wavefront shaping, based on all-optical feedback inside a laser cavity[10]. Unfortunately, it is inherently limited to point-like ("single pixel") imagery and suffers from low SNR.

In this work, we present an all-optical method, based on a highly multimode laser cavity, for *full-field* imaging through thin scattering media in real time. We show that when the numerical aperture (NA) of the laser cavity is larger than the diffusive angle of the scattering media, the laser selects the modes with minimal round-trip loss, which allow full-field imaging of the object. Due to the exponential buildup of the lasing modes[11,12], the time required for the laser to shape the beam traveling inside the cavity can be extremely short, on the order of several round trips. Specifically, we show that the buildup time can be as short as 200 ns in a Q-switched laser, orders of magnitude faster than any other reported wavefront shaping methods[13,14].

**Results**

**Concept.** The physical mechanism which enables imaging through scattering media inside a multimode laser cavity is inherently related to the process of mode build-up in lasers. In the pre-lasing stage, photons are spontaneously emitted from the gain medium and due to the random nature of the emission, a large ensemble of possible states (modes) is randomly sampled. These initial modes compete over available gain, until only modes with minimal loss, and their coherent superposition, survive in the steady state lasing stage[11,12]. Since a self-consistent laser mode must accurately retrace its path after a round-trip in the laser cavity, the minimal-loss modes possess the intrinsic property of efficient self-imaging through the intra-cavity scattering media. When many such degenerate low-loss modes exist, their coherent superposition can well approximate an arbitrary image that would then be self-imaged inside the laser cavity. Typical multimode laser cavities support only a small number of lasing modes (<100), and consequently are inadequate for optical imaging. Therefore, in our method we resort to a *highly* multimode laser that supports >100,000 spatial modes, often referred to as a degenerate cavity laser or a self-imaging cavity laser[11,15,16].

**Experimental setup and basic results.** Our degenerate cavity laser (DCL) is comprised of an Nd:YAG gain medium, two flat mirrors and two lenses in a 4*f* telescope, *f* being the focal length of the lenses (Fig. 1a and Supplementary Information). The 4*f* telescope assures that any field distribution is accurately imaged onto itself after a single round trip. Consequently, any transverse field distribution is an eigenmode of the cavity. Since all transverse modes supported by the cavity are loss-degenerate, they can all lase simultaneously, despite mode competition[11,15,17]. As the size of the exit pupil in a 4$f$ telescope is much larger than its diffraction limit, the DCL supports many transverse modes, typically well over 100,000[17]. The 4$f$ DCL is especially attractive, as it enables physical access to both the position space (mirror or output coupler planes) and k-space (Fourier space, midway between the lenses) components of the lasing beam[18–20].

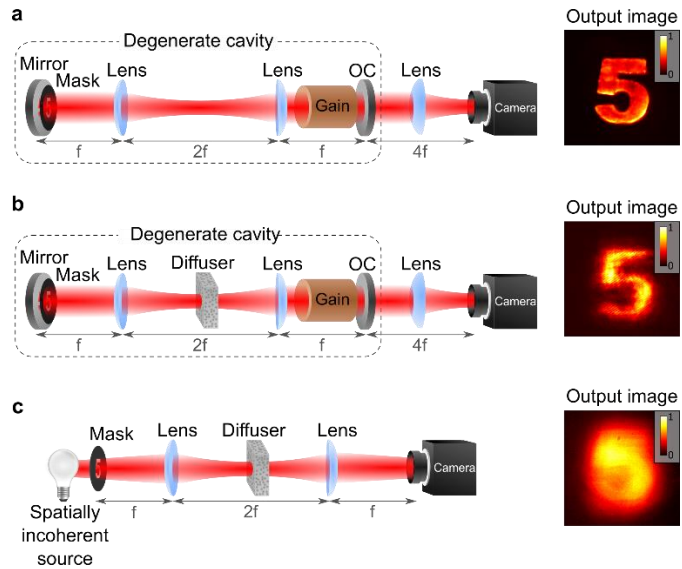

**Figure 1. Experimental arrangements for imaging through scattering media. a**, Imaging of a mask (object) in a DCL. The intensity distribution detected at the output plane (right inset), matches the shape of the object. **b**, Imaging of a mask (object) in a DCL with an intra-cavity diffuser. The DCL overcomes the effect of scattering induced by the diffuser, so the intensity distribution at the output plane (right inset) still matches the shape of the object. **c**, Imaging of a mask (object) with a passive 4$f$ telescope (no optical feedback), where a diffuser is placed at the Fourier plane of the 4$f$ telescope. Due to scattering induced by the diffuser, the detected intensity distribution (right inset) is smeared, and the object cannot be identified.

To demonstrate imaging through scattering media, a binary amplitude mask, serving as an object, is placed near the back mirror of the DCL (Fig. 1a-b). When there is no optical diffuser inside the cavity (Fig. 1a), the mask is imaged onto itself after every round trip, and therefore the lasing beam is formed in the shape of the mask[21–23] (Fig. 2, left column). Surprisingly, when the diffuser is added to the Fourier plane of the cavity (Fig. 1b), the DCL is able to overcome the effect of scattering, and the image of the object is still clearly identified at the output plane of the laser (Fig. 2, middle column). For comparison, we also placed the object masks and the diffuser in an identical passive 4*f* telescope (Fig. 1c). As expected, in this case the object cannot be identified (Fig. 2, right column).

To image through live biological tissues, it is crucial to have rapid imaging (typically, <1 µs exposure time per frame). To demonstrate such rapid imaging through the scattering media, we Q-switched the DCL by placing a Pockels cell inside the cavity. Figure 2d shows the experimental results with a Q-switched DCL where the lasing pulse duration was 200 ns. As evident, the Q-switched laser is able to image the object through the diffuser, despite the extremely short temporal duration of the lasing pulse.

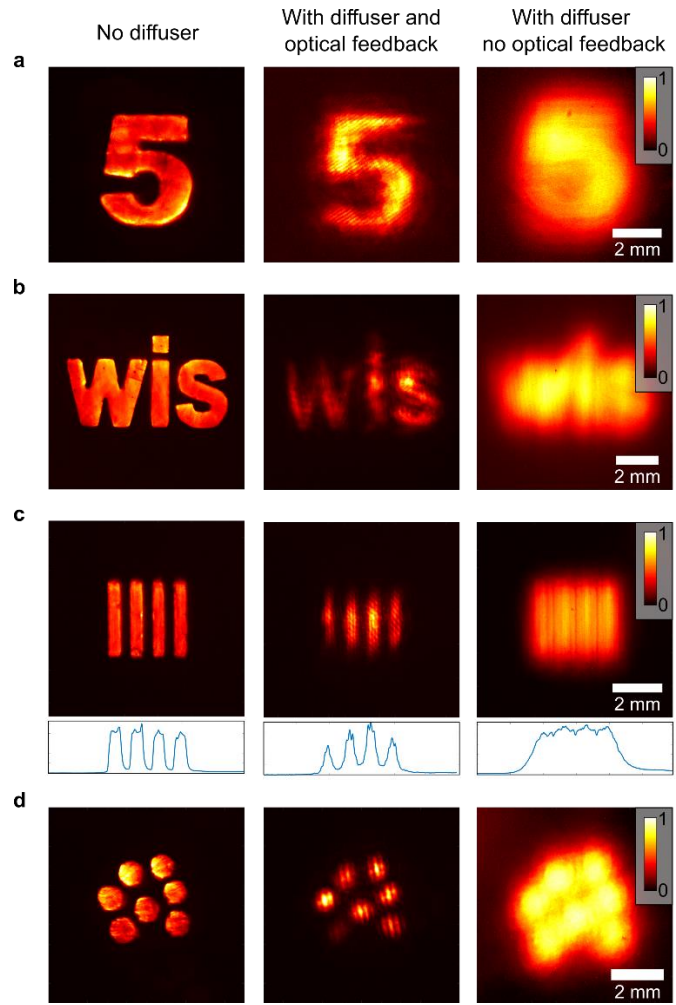

**Figure 2. Experimental results demonstrating intra-cavity imaging through scattering media for different objects. a-c**, Left column – intensity distributions of the imaged objects at the output plane of a DCL with no diffuser (as in Fig. 1a); the actual masks (objects) were placed near the back mirror inside the cavity. Middle column – intensity distributions at the output plane of a DCL, with a diffuser at the Fourier plane of the telescope (as in Fig. 1b). Right column – intensity distributions at the imaging plane of a passive 4*f* telescope (no optical feedback) with the same diffuser at the Fourier plane (as in Fig. 1c). Bottom insets in c show crossections of the intensity distributions. **d,** as in **a-c**, but here the middle column was obtained with a Q-switched cavity of 200 ns pulse duration.

**Physical mechanism.** To better understand the physical mechanism of our method, we compared the intensity distribution at the diffuser plane to the phase structure of the diffuser. For simplicity, we first considered a binary diffuser, namely a two-level optical diffuser inducing either 0 or $\pi$ phase shifts to light propagating through it (Fig. 3a, left; see Supplementary Information).

Remarkably, when the binary diffuser is placed at the Fourier plane of the DCL, lasing occurs almost entirely at uniform phase regions of the diffuser, essentially eliminating the scattering induced by the diffuser (Figs. 3a, middle and right column).The ability of the DCL to overcome the scattering induced by the diffuser requires phase locking between its many transverse modes[10,24]. Such phase locking is only possible when the detuning between modes is sufficiently small, as compared to the strength of coupling between them. Since a phase difference of $\pi$ implies maximum detuning between transverse modes, the DCL must lase only through regions of the diffuser that are of equal phase (0 or $\pi$, for the binary diffuser). Indeed, the experimental results in the right column of Fig. 3a, show that >90% of the lasing power goes through the $\pi$ phase regions of the intra-cavity diffuser, as also verified by numerical simulations, based on a Fox-Li type algorithm[11,25] (see Supplementary Information).

The fundamental limitation of our method is that the optical resolution of the intra-cavity $4f$ telescope must be high enough to focus light to the small regions of uniform phase at the plane of the diffuser. This requires that the NA of the DCL should be large compared to the divergence angle of the optical diffuser. In our experiments, we controlled the NA of the DCL by placing circular aperture objects with different diameters $D$ near its back mirror. As expected, only objects of large diameters (high NAs) were successfully imaged through the diffuser (Fig. 3b). Specifically, for objects of diameters larger than $D$=2 mm, the output lasing beam was comparable in size and shape, and lasing occurred mostly through the $\pi$ phase regions of the diffuser (Fig. 3c). However, for objects of diameters smaller than $D$=2 mm, the output beam is considerably larger than the object and lasing occurs through both 0 and π phase regions of the diffuser with approximately equal power.

The failure of the laser to overcome the effect of the diffuser for small object diameters (small NAs) is clearly manifested by the lasing intensity distribution at the plane of the diffuser (Fig. 3b, right column), and is inherently related to the modal structure of the laser cavity (see Supplementary Information). To show this more quantitatively, we calculated the autocorrelation width of the intensity distribution at the plane of the diffuser. For small object of D<2 mm, the 1/$e$ autocorrelation width is well approximated by the diffraction limit $\omega_0=0.42\lambda f/D$ of the 4$f$ telescope, which exceeds the width of typical uniform phase regions on the diffuser. For large objects of $D$>2 mm, the width of typical lasing regions is approximately constant and barely changes with object size – demonstrating the strong tendency of the DCL to select and fill only regions of uniform phase (see Supplementary Information).

Next, we placed in the intra-cavity Fourier plane a more general quasi-continuous optical diffuser, which consists of many random phase levels (see Methods). The results are presented in Fig. 4. Figure 4a shows the experimentally measured phase-structure of the diffuser (see Supplementary Information), from which the phase gradient $G$ was calculated. Figure 4b shows the overlay of the lasing intensity distribution at the plane of the diffuser and the phase gradient $G$. As evident, lasing occurs in much smaller regions than those for the binary diffuser (Fig. 3b, right column), and is limited to regions of low phase gradients, where scattering is minimized. Figure 4c shows this explicitly, by comparing the probability density distribution of the entire diffuser phase gradients $P(G)$ (dashed black) and of the diffuser regions in which the lasing intensity is above 10% of its maximal value after basic noise reduction (solid blue). Clearly, the distribution in regions with significant lasing is much narrower than that of the entire diffuser, and no lasing occurs in high phase-gradients regions of the diffuser.

Finally, we calculated the cross-correlation $C$ between phase gradients $G$ and intensity distribution $I$ at the plane of the diffuser,

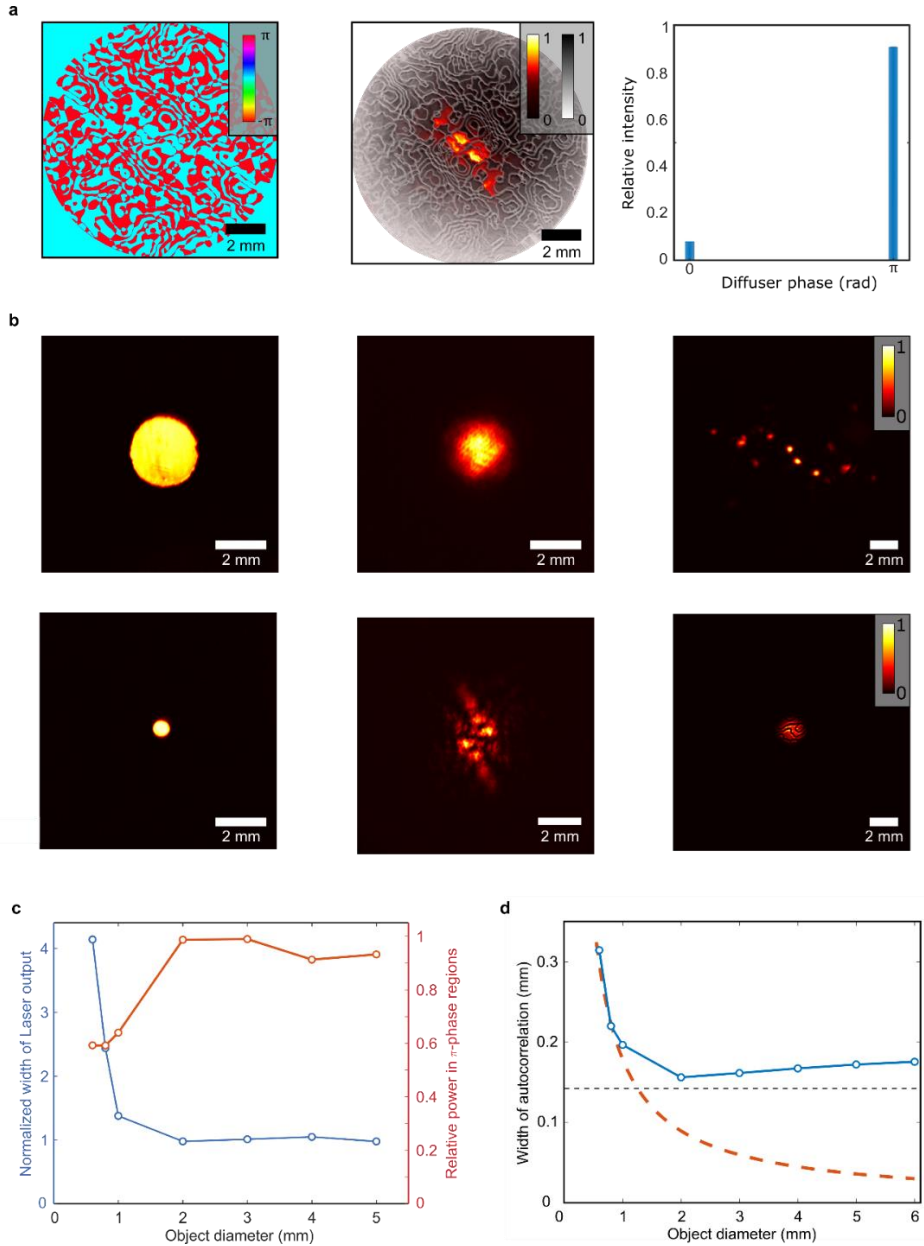

**Figure 3. Lasing through specific uniform phase regions of an intra-cavity binary diffuser. a**, Left – measured phase structure of the binary diffuser. Middle – intensity distribution at the plane of the intra-cavity binary diffuser, overlaid with the boundaries between different phase regions. Right – relative lasing power in the 0 and π phase regions of the diffuser. Lasing occurs almost exclusively through the π phase regions. **b**, Left column – intensity distributions at the output plane of the DCL with no diffuser (as in Fig. 1a), obtained with circular aperture objects of diameters *D*=2 mm (top) and *D*=0.6 mm (bottom). Middle column – intensity distributions at the output plane of the DCL, with the diffuser at the Fourier plane (as in Fig. 1b). Right column – intensity distributions at the plane of the diffuser. **c**, Solid blue – width of the output beam normalized by the width of the object, at 5% of maximal intensity, as a function of the object diameter *D*. Solid red – relative lasing power in π-phase regions of the diffuser as a function of the object diameter *D*. **d**, Autocorrelation width of the intensity distribution at the plane of the diffuser (divided by $\sqrt{2}$) as a function of object diameter (solid blue), compared with the diffraction limit $\omega_0 = 0.42\lambda f/D$ (dashed red), and the effective size of a typical uniform phase region of the diffuser (dashed black).

$$C(\Delta r) = \frac{A \int G(r)I(r+\Delta r)d^2r}{\sqrt{\int G(r)^2 d^2r \int I(r)^2 d^2r}} - 1,$$

where *r* is the position, *Δr* is a displacement in position, and $A$ is the area over which the integration is performed. We found that for *Δr*=0, the intensity distribution is anti-correlated with the phase gradient (Fig. 4d). This clearly indicates that lasing occurs only in regions of the diffuser with low phase gradients. Numerical simulations of the multi-level diffuser further validate the results (Figs. 4e-h). In particular, Fig. 4g shows that the simulated *P(G)* of the entire diffuser is much broader than that of the regions with significant lasing (as in the experimental data of Fig. 4c). For larger NA values, the DCL operation directs light to regions of low phase gradients, and consequently, the probability distribution is narrower. Figure 4h shows simulation results of the cross-correlation peak $C_0 \equiv C(\Delta r=0)$, between the diffuser phase gradients and the lasing intensity distribution, as a function of NA. As evident, for small values of NA the two are weakly correlated ($C_0 \approx 0$). However, for larger values of NA they are anti-correlated ($C_0$ approaches -1).

**Discussion**

We demonstrated a new method for rapid full-field imaging through scattering media, using a highly multimode degenerate cavity laser (DCL) and intra-cavity diffuser. We showed that this method is applicable when the optical resolution of the intra-cavity telescope is sufficiently high in relation to the correlation length (i.e. minimal feature size) of the diffuser, so as to require careful and accurate optical design and alignment. Accordingly, we demonstrated our method, as a proof of concept, with relatively weak optical diffusers. Nevertheless, we believe there is no fundamental limitation, and the method should work also with stronger optical diffusers. Furthermore, we considered here only imaging through thin scattering media; however our method should also allow finding 'open channels'[26–29], that may enable imaging through thick scattering media, when used in laser cavities with very high NAs.

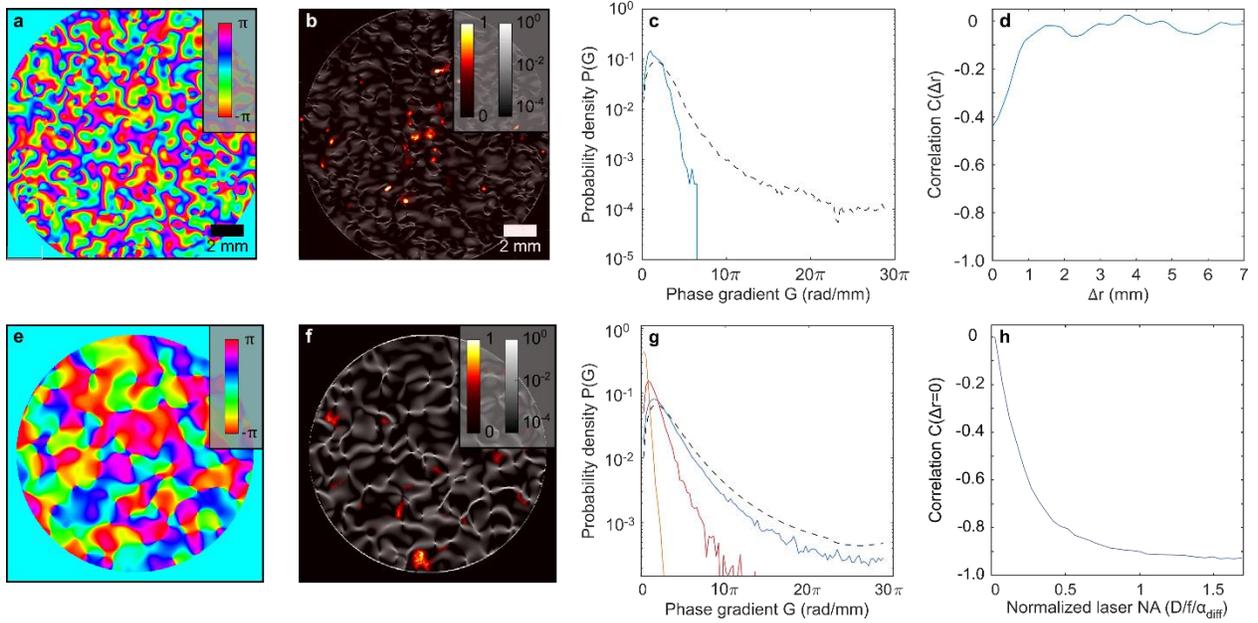

**Figure 4. Lasing through low phase gradient regions of the intra-cavity diffuser. a**, Measured phase structure of the multi-level diffuser. **b**, Absolute value of the phase gradient $G$, calculated from the phase structure in **a**, overlaid with the measured lasing intensity distribution at the plane of the diffuser. **c**, Probability distribution $P(G)$ of the entire diffuser (dashed black) and of regions of the diffuser where the lasing intensity is >10% of its maximal value (solid blue). **d**, Cross-correlation $C(\Delta r)$ between the intensity distribution at the plane of the diffuser and the phase gradient $G(\mathbf{r})$. **e**, Phase structure of the simulated multi-level diffuser. **f**, Absolute value of the phase gradient $G$, calculated from the phase structure in **e**, overlaid with the simulated lasing intensity distribution at the plane of the diffuser. **g**, Probability distribution $P(G)$ of the entire diffuser (dashed black) and of regions of the diffuser where the simulated lasing intensity is >10% of its maximal value, for different NAs of the intra-cavity telescope (normalized to the diffusion angle of the diffuser): 0.8 (solid blue), 2 (red) and 6 (solid yellow). **h**, Cross-correlation peak $C_0$ as a function of the normalized NA of the intra-cavity telescope.

Finally, we note that in recent years there has been growing interest in exploiting advanced laser systems for solving minimization problems, that occur for example in XY spin systems[30,31], Ising machine[32–34] and phase retrieval[35]. To a great extent, our work is part of this increasing effort to exploit the basic physical mechanism of lasing to solve hard problems, and may lead to novel computational and photonic applications.

## Methods

Full description of the experimental arrangements can be found in the Supplementary Information. In brief, the DCL in our experimental arrangement contained a Nd:YAG rod gain medium (wavelength $\lambda$=1064 nm) of 9.5 mm diameter and 109 mm length that was pumped with long 100 μs quasi-CW pulses of a Xenon flash lamp, at a slow repetition rate of 1 Hz, so as to minimize thermal lensing. The two lenses inside the cavity were AR coated plano-convex lenses, with diameters of $D$=50.8 mm and focal lengths of 400 mm. The back mirror was a highly reflective mirror (reflectivity $R$>99.8%), and the output coupler was a partially reflective mirror ($R$=90%). All images were obtained using a standard triggered CMOS camera (Ximea MQ013MG-E2).

Two different optical diffusers were used. Figures 2a and 4a-d were obtained with a 16-level custom made HoloOr DOE optical diffuser having a divergence angle of 0.2 degrees. Figures 2b-d and 3 were obtained with a binary custom made HoloOr DOE diffuser having a divergence angle of 0.3 degrees.

## Acknowledgements

The authors acknowledge the Israel Science Foundation (ISF) (Grant No. 1881/17) for its support.

## Author contributions

All authors contributed significantly to this work. R.C and S.M. performed the experiments and data analysis. R.C. performed the numeric simulations. N.D. and A.F. supervised the work, and C.T. provided consultation. R.C. was the leading writer of the paper, with significant contribution from all authors.

## Competing interests' statement

The authors declare no competing interests.

**References**

1. Horstmeyer, R., Ruan, H. & Yang, C. Guidestar-assisted wavefront-shaping methods for focusing light into biological tissue. *Nat. Photonics* **9**, 563–571 (2015).

2. Rotter, S. & Gigan, S. Light fields in complex media: Mesoscopic scattering meets wave control. *Rev. Mod. Phys.* **89**, 15005 (2017).

3. Bertolotti, J. *et al.* Non-invasive imaging through opaque scattering layers. *Nature* **491**, 232 (2012).

4. Katz, O., Heidmann, P., Fink, M. & Gigan, S. Non-invasive real-time imaging through scattering layers and around corners via speckle correlations. *Nat. Photonics* **8**, 784–790 (2014).

5. Edrei, E. & Scarcelli, G. Optical imaging through dynamic turbid media using the Fourier-domain shower-curtain effect. *Optica* **3**, 71–74 (2016).

6. Wang, Z., Jin, X. & Dai, Q. Non-invasive imaging through strongly scattering media based on speckle pattern estimation and deconvolution. *Sci. Rep.* **8**, 9088 (2018).

7. Vellekoop, I. M. & Mosk, A. P. Focusing coherent light through opaque strongly scattering media. *Opt. Lett.* **32**, 2309–2311 (2007).

8. Katz, O., Small, E. & Silberberg, Y. Looking around corners and through thin turbid layers in real time with scattered incoherent light. *Nat. Photonics* **6**, 549–553 (2012).

9. Hsu, C. W., Liew, S. F., Goetschy, A., Cao, H. & Douglas Stone, A. Correlation-enhanced control of wave focusing in disordered media. *Nat. Phys.* **13**, 497–502 (2017).

10. Nixon, M. *et al.* Real-time wavefront shaping through scattering media by all-optical feedback. *Nat. Photonics* **7**, 919–924 (2013).

11. Siegman, A. E. *Lasers*. (University Science Books, 1986).

12. Yariv, A. & Yeh, P. *Photonics: optical electronics in modern communications*. **6**, (Oxford university press New York, 2007).

13. Tzang, O. *et al.* 1D to 2D modulation for ultra fast focusing through complex media. in *Computational Optical Sensing and Imaging* JTh3D-1 (Optical Society of America, 2019).

14. Feldkhun, D., Tzang, O., Wagner, K. H. & Piestun, R. Focusing and scanning through scattering media in microseconds. *Optica* **6**, 72–75 (2019).

15. Arnaud, J. A. Degenerate Optical Cavities. *Appl. Opt.* **8**, 189 (1969).

16. Cao, H., Chriki, R., Bittner, S., Friesem, A. A. & Davidson, N. Complex lasers with controllable coherence. *Nat. Rev. Phys.* **1** (2019).

17. Nixon, M., Redding, B., Friesem, a a, Cao, H. & Davidson, N. Efficient method for controlling the spatial coherence of a laser. *Opt. Lett.* **38**, 3858–61 (2013).

18. Barthelemy, A. *et al.* Intracavity coherent shaping of laser beams. in *Lasers and Electro-Optics Society Annual Meeting, 1995. 8th Annual Meeting Conference Proceedings, Volume 1., IEEE* **2**, 192–193 (IEEE).

19. Chriki, R. *et al.* Manipulating the spatial coherence of a laser source. *Opt. Express* **23**, 12989 (2015).

20. Chriki, R. *et al.* Rapid and efficient formation of propagation invariant shaped laser beams. *Opt. Express* **26**, (2018).

21. Hardy, W. A. Active image formation in lasers. *IBM J. Res. Dev.* **9**, 31–46 (1965).

22. Couderc, V., Guy, O., Barthelemy, A., Froehly, C. & Louradour, F. Self-optimized resonator for optical pumping of solid-state lasers. *Opt. Lett.* **19**, 1134–1136 (1994).

23. Gigan, S., Lopez, L., Treps, N., Maaetre, A. & Fabre, C. Image transmission through a stable paraxial cavity. *Phys. Rev. A* **72**, 023804 (2005).

24. Wu, T., Chang, W., Galvanauskas, A. & Winful, H. G. Model for passive coherent beam combining in fiber laser arrays. *Opt. Express* **17**, 19509–19518 (2009).

25. Fox, a. G. & Li, T. L. T. Modes in a maser interferometer with curved and tilted mirrors. *Proc. IEEE*

**51**, 80–89 (1963).

26. Freund, I., Rosenbluh, M. & Feng, S. Memory effects in propagation of optical waves through disordered media. *Phys. Rev. Lett.* **61**, 2328–2331 (1988).

27. Imry, Y. Active transmission channels and universal conductance fluctuations. *Europhysics Lett.* **1**, 249 (1986).

28. Vellekoop, I. M. & Mosk, A. P. Universal optimal transmission of light through disordered materials. *Phys. Rev. Lett.* **101**, 120601 (2008).

29. Sarma, R., Yamilov, A. G., Petrenko, S., Bromberg, Y. & Cao, H. Control of energy density inside a disordered medium by coupling to open or closed channels. *Phys. Rev. Lett.* **117**, 86803 (2016).

30. Nixon, M., Ronen, E., Friesem, A. a. & Davidson, N. Observing geometric frustration with thousands of coupled lasers. *Phys. Rev. Lett.* **110**, 1–5 (2013).

31. Takeda, Y. *et al.* Boltzmann sampling for an XY model using a non-degenerate optical parametric oscillator network. *Quantum Sci. Technol.* **3**, 14004 (2017).

32. Inagaki, T. *et al.* Large-scale Ising spin network based on degenerate optical parametric oscillators. *Nat. Photonics* **2**, 1–3 (2016).

33. McMahon, P. L. *et al.* A fully-programmable 100-spin coherent Ising machine with all-to-all connections. *Science* **354,** *614*-617 (2016).

34. Okawachi, Y. *et al.* Coupled Degenerate Parametric Oscillators Towards Photonic Coherent Ising Machine. in *CLEO: QELS_Fundamental Science* FM1D-6 (Optical Society of America, 2019).

35. Tradonsky, C. *et al.* Rapid laser solver for the phase retrieval problem. *Sci. Adv.* **5**, eaax4530 (2019).

# Supplementary Information:

## Real-time full-field imaging through scattering media by all-optical feedback

Ronen Chriki[1], Simon Mahler[1], Chene Tradonsky[1], Asher A. Friesem[1] and Nir Davidson[1]

[1] Department of Physics of Complex Systems, Weizmann Institute of Science, Rehovot 7610001, Israel

### I. Detailed experimental arrangements

*a. Imaging through intra-cavity scattering media with a degenerate cavity laser*

Figure S1 shows schematically the typical experimental arrangement of the degenerate cavity laser (DCL) with scattering media (intra-cavity diffuser) and the external detection configuration. The DCL is comprised of an Nd:YAG gain medium rod of 109 mm length and 9.5 mm diameter, two plano-convex lenses of focal length $f$=400 mm and diameter $D$=50.8 mm in a $4f$ telescope arrangement, a highly reflective back mirror ($R$=99.5%) and a partially reflective output coupler ($R$=90%). In order to prevent issues of thermal lensing, the gain medium was pumped by a Xenon flash lamp with quasi-CW pulses of 100 µs duration at slow repetition rate of 1 Hz. The object was a binary amplitude mask placed adjacent to the back mirror, and an optical diffuser (which served as the scattering media) was placed at the Fourier plane between the two lenses of the $4f$ telescope. In our work we used two different optical diffusers: a 16-level custom made HoloOr DOE optical diffuser having a divergence angle of 0.2 degrees, and a binary custom made HoloOr DOE diffuser having a divergence angle of 0.3 degrees.

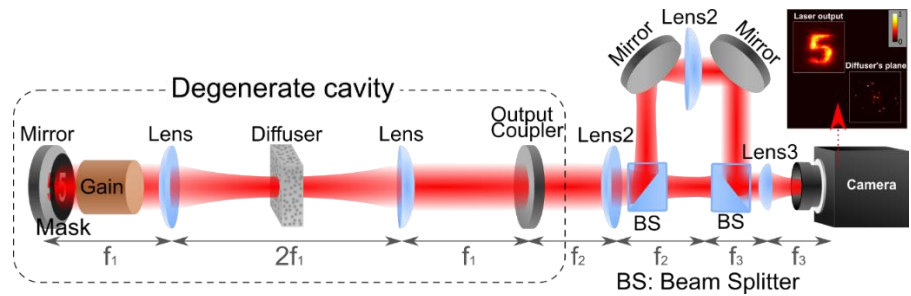

**Figure S1.** Experimental arrangement of the DCL with imaging system, used for measuring the intensity distributions at the laser output and at the diffuser's plane.

The telescope assured perfect imaging in the laser cavity whereby any field distribution is precisely imaged onto itself after a single round trip, and therefore any field distribution is an eigenmode of the cavity. Typically, the size of the exit pupil in the $4f$ telescope is much larger than the diffraction limit, so many transverse modes can all lase simultaneously. We noticed experimentally that the results are essentially the same when the gain medium is placed either near the back mirror or near the output coupler (not shown).

A beam splitter placed outside the cavity splits the laser's output to two arms. The first arm images the intensity distribution at the output plane of the laser onto a camera (by means of Lens2 and Lens3), and the second arm images the intensity distribution at the diffuser plane (by means of the intra-cavity Lens1, Lenses 2 and Lens3) onto the same camera, but with a transverse shift as compared to the first arm; accordingly, the light from the two arms illuminate different regions of the camera. The camera therefore detects simultaneously the intensity distributions at both the output plane and the diffuser plane. The focal distance of Lenses2 was 200 mm and their diameters were 25.4 mm. The focal distance of Lens3 was 60 mm and its diameter was 25.4 mm.

### b. *Basic imaging through scattering media with a spatially incoherent source*

Figure 1c in the manuscript shows schematically the basic experimental arrangement for imaging through scattering media outside a laser cavity with a spatially incoherent source, which was used to obtain the results of Figs. 2 (right column). The spatially incoherent source was obtained by illuminating a fast rotating diffuser with a single mode continuous wave Nd:YAG laser. The light from this source propagated through the object (a binary amplitude mask) and the output intensity distribution was then imaged through the scattering media and onto a CMOS camera using a 4$f$ telescope configuration (the focal lengths of the telescope lenses were $f$=400 mm and their diameters $D$=50.8 mm). The scattering media were optical diffusers placed midway between the two telescope lenses, at the Fourier plane.

### II. Phase-structure measurements of the optical diffuser

Figure S2 schematically presents the experimental arrangement and the Fourier processing procedure for measuring the phase-structure of the optical diffuser (e.g. as in Fig. 4a). The experimental arrangement shown in Fig. S2a is a Mach-Zehnder interferometer whose input was a relatively large diameter collimated light beam from a single mode continuous wave 1064 nm laser. The input laser light beam was divided by the first beam splitter into two light beams of equal intensities, so each propagates in a different channel. In the first channel, the light beam propagates through a telescope, formed by two spherical lenses of focal lengths $f$=200 mm and diameter $D$=50.8 mm, and the diffuser, at the focal plane of the lenses, to the second beam splitter. In the second channel, the light beam directly propagates at an angular orientation to the second beam splitter. At the second beam splitter plane, the two light beams interfere, forming interference fringes that are imaged to and recorded by a camera, as in Fig. S2b (right image).

Figure S2b shows the Fourier processing procedure for determining the phase-structure of the optical diffuser from the intensity distribution of the interference fringes (right image). First, the intensity distribution of the detected interference fringes was Fourier transformed. Then an aperture was applied in Fourier space to select one of the two high frequency lobes, and move it to the center. Finally, an inverse Fourier transform is applied to obtain the phase distribution at the plane of the diffuser, and thereby the phase-structure of the optical diffuser (left image). To compensate for aberrations and improve accuracy, we repeated the procedure after removing the diffuser, and subtracted the resulting phase distribution at the plane of the diffuser from that with the diffuser.

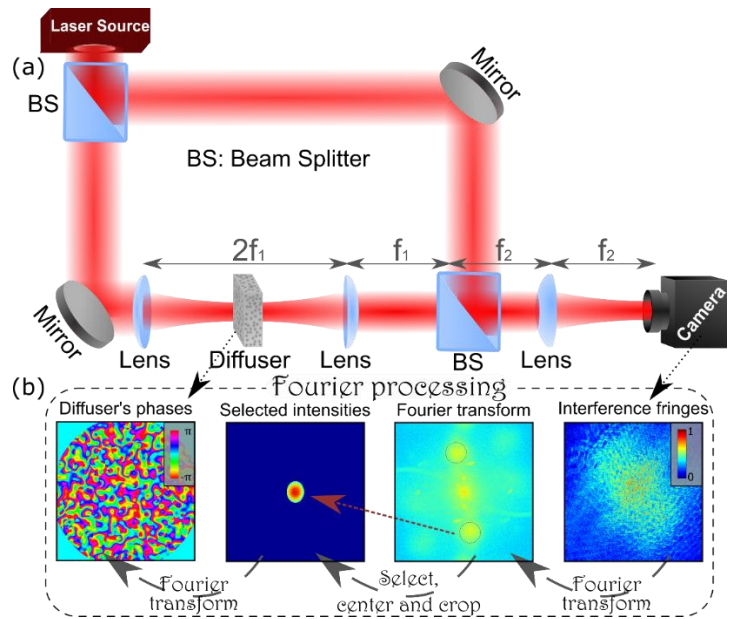

**Figure S2.** Experimental arrangement and Fourier processing method used for measuring the phase-structure of the optical diffuser.

For the case of the binary diffuser, we compared the results of this method with those obtained by imaging the intensity distribution at the diffuser plane directly onto the camera, using a spatially incoherent source. As the binary diffuser is comprised of regions that induce 0 and π phase shifts, destructive interference occurs at the boundary between these regions, as seen in the middle column of Fig. 3a. Therefore, the different phase regions can easily be identified. We found that the results from this direct imaging method are in very good agreement with those from the more elaborate Fourier processing method of Fig, S2b. Since it is simpler, cleaner and less noisy, we mainly used the direct imaging method throughout our experiments.

It should be noted that when the diffuser is inside the DCL (see Fig. S1), light propagates through the diffuser twice every round trip, i.e. forward and backward directions. Due to the inversion property of a 4*f* telescope, the field of one direction is rotated by 180° relative to the other direction. This 180° rotation effect was numerically taken into account in our measurements and analysis.

**III. Simulation method**

The simulations presented in the main text are based on a Fox-Lee type simulation[1,2], where the minimum loss eigenmode of the cavity is found by iteratively propagating a transverse field through the cavity, until a steady state is finally achieved. The transverse field is represented by a field matrix of NxN elements (pixels), and it is initially generated by distributing random phases and uniform intensity over all matrix elements. The optical components and propagation distances in the cavity are represented by a matrix, and therefore a single round trip in the cavity is represented by matrix multiplication. For the DCL, the propagation from the plane of one of the mirrors to the diffuser, and vice versa, is represented by a fast Fourier transform. After each round trip, the field is normalized, so as to account for loss and gain and avoid numerical instabilities.

Figure S3 compares simulated intensity distributions to those measured experimentally. For the simulations the shape of the object is the same as the physical mask that was placed inside the DCL, and the diffuser has the phase structure of the binary diffuser (which we measured by the direct imaging method). The simulated results were obtained by summing the intensity distributions of 300 different simulations, to account for the effect of different longitudinal modes, which are assumed to be uncorrelated with one another. The number of longitudinal modes in the cavity is estimated by the measured bandwidth of our laser (measured to be ~30GHz, not shown) divided by the spectral spacing between modes ($c/8f$=9.4MHz, where *c* is the speed of light, and *f* is the focal length of the lenses inside the cavity).

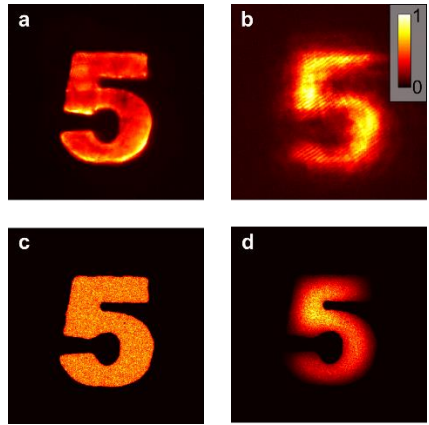

**Figure S3. Comparison between simulation and experimental results. a, c,** Experimental and simulated intensity distributions at the output plane of the DCL, when there is no diffuser inside the cavity (see Fig. 1a in the manuscript). **b, d,** Experimental and simulated intensity distributions at the output plane of the DCL, when there is a diffuser inside the cavity (see Fig. 1b in the main text).

Similarly, the results presented in Fig. 4f in the main text were obtained by summing the intensities of 100 individual simulations, where each simulation result was obtained after 200 iterations. The results of Fig. 4g in the main text were achieved by running 100 separate simulations, each for a different optical diffuser but with similar properties, i.e. all have the same size and the same divergence angle. Each simulation in Fig. 4g was obtained after 1000 iterations. The results of Fig. 4h were achieved by running 20 separate simulations, each for a different optical diffuser but with similar properties. Each simulation was obtained after 4000 iterations.

As mentioned in the main text, the optical diffuser couples between the different spatial modes, such that they interact with one another and tend to find a constant relative phase that will minimize loss. Typically, the minimal loss mode is self-consistent, and therefore allows imaging of the object through scattering media (diffuser). However, the diffuser also induces random variations in optical length, and thus different spatial modes will lase with different frequencies. This detuning in lasing frequencies causes round trip loss, and therefore limits the ability of different transverse modes to lock in phase[3,4]. If the coupling between modes is sufficiently strong, as compared to the degree of frequency detuning between modes, the gain of the phase-locked state would be greater than the loss of the non-phase-locked state, and therefore the coupled modes will find a mutual lasing frequency in a process of frequency pulling. However, if the detuning is large as compared the coupling, the modes will not phase lock to one another.

In order to account for this effect in our simulations, we added a loss matrix which mimics the effect of frequency detuning. For this purpose, we span all modes in the DCL using a basis where each mode is a plane wave with different angular orientation near the mirrors, and a different diffraction limited spot at the Fourier plane between the two lenses. Accordingly, each pixel (*i*, *j*) in the plane of the diffuser represents an individual mode. In a ring cavity, where light propagates through the diffuser only ounce every round trip, and in the absence of coupling, the relative phase $\varphi_{i,j}^0$ between a single mode and its overlap after a round trip is just

$$\varphi_{i,j}^0 = \frac{2L\omega_0}{c} + \theta_{i,j},$$

where *L*=4*f* is the length of the cavity, *c* is the speed of light, $\theta_{i,j}$ is the phase induced by the optical diffuser at pixel (*i*, *j*) and $\omega_0$=$\pi nc/L$ is some natural frequency of the DCL in the absence of the diffuser with *n* being an integer number. The lasing frequency of the uncoupled mode is then $\omega_{i,j} = c\varphi_{i,j}^0/2L$, and the phase accumulated by the coupled mode after a single round trip $\varphi_{i,j}$ contributes a round-trip loss of

$$L_{i,j} = \frac{1}{2}\left[1 - \cos(\varphi_{i,j} - \varphi_{i,j}^0)\right]$$

for each mode (*i*, *j*). By multiplying the field matrix by this loss matrix at every round trip, we are able to account for the effects of frequency detuning caused by the optical diffuser.

Figure S4 compares the simulated results with and without taking the loss matrix into account, for the two types of diffusers that were used – the binary diffuser and the multi-level diffuser. Juxtaposing these results to those of Fig. 3b and 4b in the main text indicates that the simulation results quantitatively agree with experimental results only when this loss mechanism due to frequency detuning is taken into account.

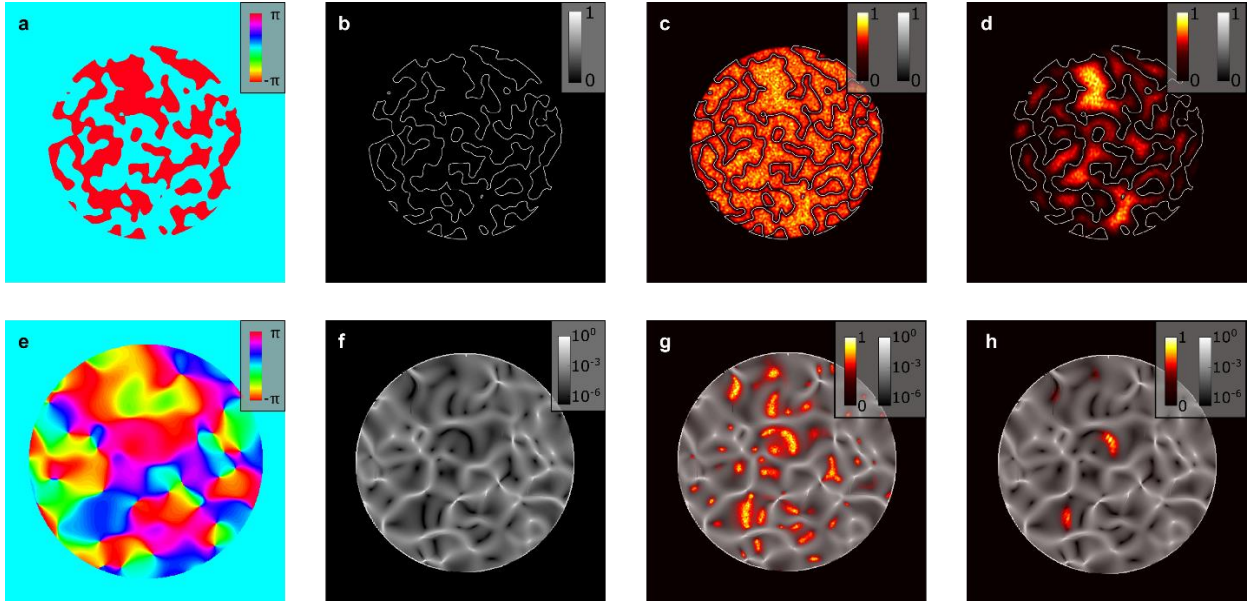

**Figure S4. Simulation results with and without taking the frequency detuning induced loss into account. a-d,** binary diffuser; **e-h,** multi-level diffuser. **a,e,** Phase structure of the simulated diffuser. **b,f,** Normalized phase gradient of the simulated diffuser. **c,g,** Overlay of the phase gradient of b and f and the simulated intensity distribution at the output of the DCL, when the effect of frequency detuning induced loss is *not* considered. **d,h,** Overlay of the phase gradient of b and f and the simulated intensity distribution at the output of the DCL, when the effect of frequency detuning induced loss *is* considered.

## IV. Modal structure

The ability of the DCL to image through scattering media is strongly related to the modal structure of the cavity, and a thorough understanding of such imaging can be obtained by analyzing the modal structure. For this purpose, consider a two dimensional DCL and a field vector $E_{in}$ of *n* elements, which represents the field $E(x; z_0)$ at some initial plane $z_0$ inside the cavity. Each propagation distance and each optical element is represented in matrix form, and therefore a single round trip is represented by a single matrix *M*. Thus, the field $E_{RT}$ after a single round trip is readily calculated,

$$E_{RT} = M \cdot E_{in}.$$

If steady-state self-consistent modes are assumed, $E_{RT} = E_{in}$, the equation above reduces to an eigenvalue problem, where the eigenvectors represent eigenmodes of the cavity and the eigenvalues are related to roundtrip loss.

We simulate a DCL with an aperture of varying sizes 2*D* near the back mirror and an aperture of size 2d in the Fourier plane between the two intra-cavity lenses. Figure S5a shows the calculated loss as a function aperture size (Fresnel number) for the lower order modes, without an intra-cavity optical diffuser. As evident, the modal structure is well ordered. As the size of the aperture is increased, there is a sharp decrease from high loss to zero for each mode. For large apertures, many modes have zero loss and are therefore loss degenerate. As a result, many modes can lase simultaneously, despite mode competition. The number *N* of loss-degenerate modes (i.e. the number of expected lasing modes) depends on the ratio between the NA of the intra-cavity 4*f* system (*D/f*) and the divergence angle of the aperture at the Fourier plane ($\lambda/\pi d$), with 2d the size of the aperture and λ the wavelength, namely

$$N = \frac{\pi d D}{\lambda f} = \frac{D}{w_0},$$

where $w_0 = \lambda f/\pi d$ is the size of a diffraction limited spot at the output plane of the DCL. In three dimensions, $N = N_x \cdot N_y \sim (D/w_0)^2$.

Figure S5b presents simulation results of loss as a function of aperture size, with an optical diffuser inside the cavity. For small aperture sizes, the NA of the cavity is small compared to the divergence angle of the diffuser, and therefore the modes are highly lossy. As a result, imaging through scattering media cannot be achieved. However, for large apertures, the NA of the cavity becomes large compared to the divergence angle of the diffuser, and therefore there are modes with low levels of loss. For sufficiently large apertures (i.e. sufficiently large NA), the loss of many modes is close to zero, and therefore many modes can lase simultaneously. The fact that the loss goes to zero implies that the modes are efficiently self-consistent, and therefore can be used for imaging through the scattering media. Due to mode competition, modes with higher loss levels are not expected to lase. The number $N_{diff}$ of loss-degenerate modes is the ratio between the NA of the cavity and the divergence angle $\alpha_{diff}$ of the diffuser

$$N_{diff} = \frac{NA}{\alpha_{diff}} = \frac{\pi D a}{\lambda f} = \frac{D}{D_0},$$

where 2*a* is a typical size of a single-phase region on the diffuser, and $D_0 = \lambda f/\pi a$. In 3D, $N_{diff} = N_x^{diff} \cdot N_y^{diff} \sim (D/D_0)^2$.

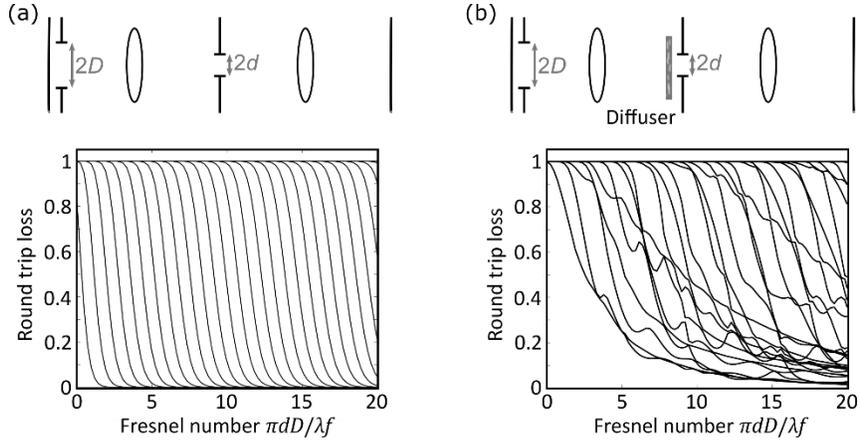

**Figure S5.** Simulated modal structures of a degenerate cavity laser, showing round-trip loss as a function of Fresnel number. (a) Without an intra-cavity diffuser, and (b) with an intra-cavity diffuser at the Fourier plane between the two lenses.

In order to verify the simulation results, we placed circular apertures near the back mirror of the DCL and measured the number of lasing modes as a function of aperture size, when a binary optical diffuser was placed at the Fourier plane between the two lenses inside the cavity. The number of modes was estimated by placing a second optical diffuser (Newport light shaping diffuser, 10° diffusion angle) outside the cavity and detecting the far field intensity distribution with a CMOS camera. The number of modes $N$ was estimated by the ratio of the standard deviation of the intensity distribution $\sigma$ over the mean of the intensity distribution $\langle I \rangle$, as

$$N = \frac{\sigma}{\langle I \rangle}.$$

Figure S6 presents experimental results for the number of modes as a function of aperture area. As evident, for sufficiently large apertures, the number of modes grows linearly with the aperture area, in agreement with our conclusions above. Moreover, the slope of the linear region is 0.91 mm$^{-2}$, which is close to the expected slope of 0.97 mm$^{-2}$.

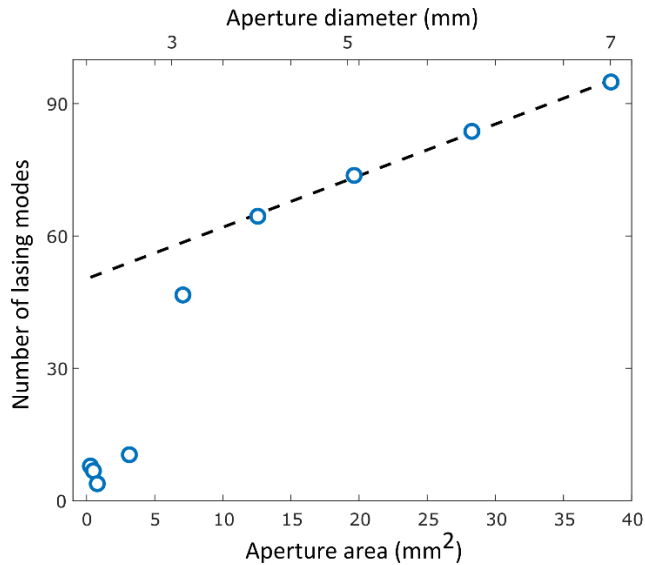

**Figure S6.** Experimental measurements of the number of lasing modes as a function of aperture area (blue circles), together with a linear fit of the large aperture data points (dashed black line).

**References**

1. Fox, a G. & Li, T. Resonant Modes in a Maser Interferometer. *Bell System Technical Journal* **40**, 453–488 (1961).

2. Siegman, A. E. *Lasers*. (University Science Books, 1986).

3. Fabiny, L., Colet, P., Roy, R. & Lenstra, D. Coherence and phase dynamics of spatially coupled solid-state lasers. *Phys. Rev. A* **47**, 4287–4296 (1993).

4. Fridman, M., Nixon, M., Ronen, E., Friesem, A. A. & Davidson, N. Phase locking of two coupled lasers with many longitudinal modes. *Opt. Lett.* **35**, 526–528 (2010).

5. Nixon, M., Redding, B., Friesem, a a, Cao, H. & Davidson, N. Efficient method for controlling the spatial coherence of a laser. *Opt. Lett.* **38**, 3858–61 (2013).